\documentclass[showkeys,nofootinbib,amsmath,
  amssymb]{revtex4-2}

\usepackage[utf8]{inputenc} 
\usepackage[T1]{fontenc} 
\usepackage[english]{babel} 
\usepackage{xcolor} 
\usepackage{amsfonts,amsthm,braket,dsfont,comment,leftindex,bbm} 
\usepackage{graphicx} 
\usepackage{caption}
\usepackage{subcaption,ragged2e} 
\usepackage{tikz} 
\usetikzlibrary{arrows.meta}
\usepackage{multirow,afterpage,tabularray} 
\usepackage{fancyhdr,orcidlink} 
\usepackage[capitalise]{cleveref} 
\usepackage{enumitem} 
\usepackage{titlesec} 
\usepackage{lipsum}
\usepackage{listings} 
\usepackage{cancel}
\usepackage{stmaryrd}
\usepackage{booktabs}
\usepackage{youngtab}

\newcommand{\diff}{\ensuremath{\mathrm{d}}}
\newcommand{\bal}{\begin{aligned}}
\newcommand{\eal}{\end{aligned}}

\definecolor{ao(english)}{rgb}{0.0, 0.5, 0.0}

\renewcommand{\bm}[1]{\boldsymbol{#1}}

\titleformat{\subsubsection}
   {\normalfont\fontsize{9pt}{11pt}\selectfont\bfseries}
   {\thesubsubsection.}
   {1em}
   {\centering}
\titleformat*{\subsubsection}{\normalsize\itshape}

\begin{document}

\title{Light hybrid baryons in the constituent model of QCD}

\author{\surname{Joachim} Viseur \orcidlink{0009-0006-3348-4402}}
\email[E-mail: ]{joachim.viseur@umons.ac.be}
\affiliation{Service de Physique Nucl\'{e}aire et Subnucl\'{e}aire,
Universit\'{e} de Mons,
UMONS Research Institute for Complex Systems,
Place du Parc 20, 7000 Mons, Belgium}

\author{\surname{Claude} Semay \orcidlink{}}
\email[E-mail: ]{claude.semay@umons.ac.be}
\affiliation{Service de Physique Nucl\'{e}aire et Subnucl\'{e}aire,
Universit\'{e} de Mons,
UMONS Research Institute for Complex Systems,
Place du Parc 20, 7000 Mons, Belgium}

\author{\surname{Cyrille} Chevalier \orcidlink{}}
\email[E-mail: ]{cyrille.chevalier@umons.ac.be}
\affiliation{Service de Physique Nucl\'{e}aire et Subnucl\'{e}aire,
Universit\'{e} de Mons,
UMONS Research Institute for Complex Systems,
Place du Parc 20, 7000 Mons, Belgium}

\date{\today}

\begin{abstract}
Hybrid baryons, in which gluonic degrees of freedom play an explicit dynamical role, provide a key testing ground for nonperturbative quantum chromodynamics. In this work, we investigate the mass spectrum of light hybrid baryons composed of identical quarks within a phenomenological constituent framework, applied to a quark core-gluon approximation. In this approach, the hybrid baryon is described as a bound state of a color-octet three-quark core and a constituent gluon, allowing the original four-body problem to be reduced to a three-body calculation followed by an effective two-body treatment. The spectrum of the color-octet quark core is obtained by solving a semirelativistic three-quark Hamiltonian with linear confinement, Coulomb, and regularized hyperfine interactions using an oscillator basis expansion. Finite-size effects of the core are incorporated through the convolution of the effective core-gluon interaction with the spatial quark density. The resulting two-body problem, whose associated Hamiltonian has the same shape as the one of the core, is solved applying the helicity formalism and using the Lagrange mesh method. Our results predict the lightest hybrid baryons to occur at energies above $3~\mathrm{GeV}$, with negative-parity states generally lying below their positive-parity counterparts. The predicted spectra are compared with lattice QCD and QCD sum-rule calculations, showing qualitative agreement although the lowest-lying lattice QCD results are significantly lighter than the present ones. Possible extensions of the model and implications for future experimental searches are discussed. 
\end{abstract}

\maketitle


\section{Introduction}
\label{sec:intro}

The identification and characterization of hybrid hadrons, i.e.\ states in which gluonic degrees of freedom play an active dynamical role, constitute important challenges in contemporary hadron spectroscopy. While quantum chromodynamics (QCD) predicts the existence of nonconventional baryons with explicit gluonic excitations~\cite{Meye15,Wang26,Dudek12,Caps02,Simo99}, their experimental confirmation remains elusive. In recent years, several experimental programs have significantly strengthened the motivation for a systematic study of hybrid baryons. The GlueX Collaboration at Jefferson Lab has initiated dedicated searches for hybrid baryons through photoproduction channels. However, to date, the most compelling experimental indications of hybrid states have been observed only in the meson sector~\cite{GlueX21,Hurc25}. Besides, analyses from CLAS12 suggest that certain excited baryons in the light-quark sector may accommodate gluonic excitations, although current data do not yet allow a clear discrimination from conventional three-quark configurations~\cite{Moke18}.

From the theoretical point of view, hybrid baryons have been studied using a wide variety of nonperturbative approaches. Lattice QCD (LQCD) simulations predict the existence of low-lying hybrid baryons with masses in the range $2.5-3$~GeV, featuring characteristic gluonic excitations with quantum numbers not compatible with simple three-quark configurations~\cite{Dudek12}. The flux-tube model~\cite{Caps02} and QCD sum rules~\cite{Kiss95,Aziz17,Wang26} also confirm the existence of gluon-rich baryonic excitations, though with significant spread in the predicted mass hierarchies and state multiplicities. In the constituent picture, hybrid baryons are treated as effective bound states of three quarks and one constituent gluon, interacting via a QCD-motivated potential. However, the direct four-body treatment remains technically involved due to the presence of gluon helicity degrees of freedom and the complexity of constructing properly symmetrized four-body helicity states.

In this work, we address these difficulties by employing the \emph{quark core–gluon model} recently introduced in Ref.~\cite{Cimi24}. Instead of solving the full four-body problem, the three light quarks are first combined into an effective \emph{quark core} with definite $J^P$ and color-octet structure. The hybrid baryon then emerges as a two-body bound state of this quark core and a constituent gluon, whose quantum numbers are taken to be the natural ones, i.e $J^P=1^-$. This picture is analogous to the quark-diquark approximation of ordinary baryons, which recently turned out to give good results in constituent framework, in both the heavy and light sectors \cite{tour25}.

Besides, the core-gluon-like construction provides technical advantages. The complexity of the Hilbert space is dramatically reduced: only two-body helicity states are required, for which the associated formalism is well established~\cite{Jaco59,chev25.1}. This effectively makes the two-body system resemble a two-gluon glueball at the color level, a picture that is well described by many QCD models~\cite{llan21,math09}. The nontrivial internal structure of the quark core is, for its part, taken into account through a convolution of the core-gluon potential with the spatial quark distribution obtained from the three-body calculation. This study mainly focuses on the prediction of mass spectrum for the lightest hybrid baryons with three identical quarks, $nnng$, where the quark core is considered in its ground state.

The paper is organized as follows.  
In Sec.~\ref{sec:quark core-gluon}, we describe the core-gluon model and its theoretical motivations.  
In Sec.~\ref{sec:quark core spectrum}, we present the three-body Hamiltonian for the quark core and compute its energy spectrum.  
Sec.~\ref{sec:q c- g interaction} constructs the effective core-gluon potential, including the convolution with the quark density.  
Sec.~\ref{sec:HB spectrum} details the construction of hybrid baryon states within the helicity formalism and the numerical resolution of the two-body problem using the Lagrange Mesh Method (LMM). Sec.~\ref{sec: spectrum} gives our hybrid baryon spectrum and presents a comparison with the literature.  
Finally, Sec.~\ref{sec: conclusion} summarizes the findings and provides perspectives for future investigations.


\section{The quark core-gluon model for light quarks}
\label{sec:quark core-gluon}

In this work, the description of hybrid baryon systems is realized using a quark-constituent model: they are considered as bound states of three quarks and a constituent gluon, interacting via a QCD-inspired potential. In this framework, a Schrödinger-like equation is to be solved to get the spectrum of the particle of interest. It is shown in \cite{math08} that considering helicity degrees of freedom instead of spin while dealing with constituent gluons leads to better agreement with LQCD calculations. But constructing four-body helicity states is a very challenging task. Even for three-body systems the construction of helicity states is cumbersome (see \cite{chev25.1} for further details). Specific approximation schemes can help mitigate this issue. The one we will use in this study, presented in \cite{Cimi24}, is called \emph{quark core-gluon model}. It consists in dividing the four-body problem in two sub-problems which are easier to deal with. Concretely, we bring the three quarks together in a \emph{quark core}, which will then interact with the remaining constituent gluon as a two-body system. In this way, we will just have to apply the well-known two-body helicity formalism~\cite{Jaco59} to construct the physical states. It is important to note that only color interactions are considered. The principle of color confinement forces the hybrid baryon to lie in a color singlet state, $\bm{1}$. Since the state associated with the gluon lies in the octet representation, $\bm{8}$, of the color symmetry group $SU(3)_c$, the three-body state associated with the quark core has to lie in that same representation, as~\cite{Boul08} 
\begin{equation}
    \boldsymbol{8} \otimes \boldsymbol{8} = \boldsymbol{1}_S \oplus \bm{8}_S \oplus \bm{8}_A \oplus \bm{10}_A \oplus \bm{\bar{10}}_A \oplus \bm{27}_S.
\end{equation}
Octet representations of the quark core stem from the following decomposition \cite{Clos79}
\begin{equation}
    \boldsymbol{3} \otimes  \boldsymbol{3}\otimes  \boldsymbol{3} = ( \boldsymbol{\bar3}_A \oplus \boldsymbol{6}_S ) \otimes  \boldsymbol{3} = (\boldsymbol{1}_A \oplus \boldsymbol{8}_{MA}) \oplus(\boldsymbol{8}_{MS} \oplus \boldsymbol{10}_S),
    \label{irrep of tensor product of 3 rep}
\end{equation}
where the indices $A,~S,~MA$ and $MS$ reflect the symmetry of the color states under exchange of the quarks. $A$ stands for \emph{antisymmetric}, $S$ for \emph{symmetric} and $M$ for \emph{mixed}.
This illustrates that, although quark cores seem very similar to ordinary baryons, they are not lying in the same color representation space. This raises the question of the way such an octet core should interact. In this study, this issue will be overcome by assuming the universality of color interactions. It means that the only difference between all the interactions described in the model is the color charge of the sources. In particular, the Hamiltonian associated with the quark-quark interactions in the quark core will have the same shape as for ordinary baryons, as well as for the core-gluon interaction in the hybrid baryons. The only differences lie in how the color part of that operator acts on the states, and the values of the parameters. Note that this hypothesis, sometimes called \emph{Casimir scaling hypothesis}, has been verified using Lattice computations for interactions similar to those used in the present work~\cite{Bali00}.

\vspace{1em}

Note that the basic idea behind the core-gluon approximation stems from the quark-diquark picture of ordinary baryons, which was the first attempt of sub-structure description of multiquarks states~\cite{Carl07,Gian19}. Although intuitively more convenient for the description of baryons containing two heavy quarks, such as $bbn$ or $ccn$~\cite{Gian09,Maje16,Farh23}, a recent study~\cite{tour25} showed that such an approximation holds for the lightest baryons as well. This encourages the use of the core-gluon model to describe not only heavy hybrids but also the lightest ones, which are the most likely to be detected.

\vspace{1em}

In view of what has been explained above, the core-gluon approximation suggests developing the following two-step procedure that leads to the determination of the mass spectrum of hybrid baryons:
\begin{itemize}
    \item Compute the energy spectrum of the color-octet quark core by solving the quantum three-body problem. It will be done by using a semirelativistic Hamiltonian with linear confinement, Coulomb interaction, and a regularized hyperfine term. The eigenstates will be obtained via an oscillator basis expansion (OBE) method. 
    \item  Solve the resulting quark core-gluon two-body problem within the helicity formalism by treating the quark core as an extended effective particle.  The interaction potential will have the same shape as for the quark core, corrected by convolution with the spatial quark density of the core to account for size effects. The mass spectrum of the light hybrid baryons will finally be  determined using the LMM, suitably adapted to handle states with helicity degrees of freedom.
\end{itemize}
We will do that step by step in the next sections.


\section{The spectrum of the quark core}
\label{sec:quark core spectrum}
The first step when using the core-gluon approach is to calculate the mass spectrum of the quark core. The obtained eigenvalues will serve as mass parameters for the core in the subsequent two-body problem (cf. Section \ref{sec:HB spectrum}), and the associated eigenstates will serve to encode the core spatial extent in the core-gluon interaction (cf. Section~\ref{sec:size effect}).
To do so, we need to solve a quantum three-body problem in its center of mass frame (CoMF). This task is tackled numerically using the OBE method. The considered Hamiltonian is the following:

\begin{equation}
    \mathcal{H}_C = \sum_{i=1}^3 \sqrt{\boldsymbol{p}_i^2+m_i^2} + \sum_{i<j=2}^3 \Big( \sigma \hspace{0.3em} r_{ij} + \boldsymbol{F}_i\cdot\boldsymbol{F}_j \hspace{0.3em} \frac{\alpha_s }{r_{ij}} + V_0 - \boldsymbol{F}_i\cdot\boldsymbol{F}_j \hspace{0.3em} \frac{8 \alpha_{ss} \Lambda^3}{3 \sqrt{\pi}m_im_j} e^{-\Lambda^2r_{ij}^2} \boldsymbol{s}_i\cdot\boldsymbol{s}_j \Big).
    \label{quark core ham}
\end{equation}
Note that $m_i$ and $\bm{p}_i$ stand for the constituent mass and the momentum of the $i$th quark respectively (with $\sum_{i=1}^3\bm{p}_i=\bm{0}$), $r_{ij} = |\boldsymbol{r}_i - \boldsymbol{r}_j|$ for the norm of the relative distance between quarks $i$ and $j$, and $\boldsymbol{F}_i$ and $\boldsymbol{s}_i$ for the color and spin operators of the $i$th quark. Inspired from~\cite{Theu01}, the Hamiltonian~\eqref{quark core ham} has the following features:
\begin{itemize}
    \item A semi-relativistic kinematic with constituent quark masses $m_i$, more appropriate for light quarks dynamics, is used.
    \item The central part is a Cornell-like potential, which contains three terms. The linear one represents the confinement of the quarks into the core, which is identical to the one for ordinary baryons~\cite{Cimi24}. The Coulomb one represents the color charge interaction, and the last one is a shift constant, justified by light hadron phenomenology~\cite{caps86}.
    \item  The last term is a hyperfine one, whose shape arises from a non-relativistic limit of the spin-spin part of the one-gluon exchange process~\cite{Math07}. However, by comparison with~\cite{Math07}, this term has here been slightly modified in two ways: first, the mass parameters $m_i$ are taken to be the same as in the kinetic energy. Second, in Ref.~\cite{Math07}, the spin-spin term lets a contact Dirac delta interaction appear. Because this would lead to divergences while solving the Schrödinger equation, in~\eqref{quark core ham}, we have chosen to regularize it, as in Ref.~\cite{Theu01,caps86}, but with a Gaussian function, which makes a new independent parameter $\Lambda$ appear.
\end{itemize}

The parameters of the model are fixed to the same values as in~\cite{Theu01}, except from those associated with the hyperfine term, i.e $\Lambda$ and $\alpha_{ss}$. In that paper, the spin-spin interaction is regularized by a Yukawa function instead of a Gaussian. A Gaussian is used in this study because it is more suitable with respect to the OBE, leading to faster convergence, hence more accurate results.  The parameters $\Lambda$ and $\alpha_{ss}$ are therefore to be refit so that the eigenvalues of~\eqref{quark core ham} agree with the experimental masses for the lightest baryons, that are the nucleon and the $\Delta$. Such a regularization gives essentially the same results as in Ref.~\cite{Theu01} for the light baryon spectrum, with deviations up to $50$~MeV maximum. Note that the parameter $\alpha_{ss}$ is actually supposed to be the same as the strong coupling constant $\alpha_s$ from the point of view of one gluon exchange diagram. However, for the same reason as for the choice of constituent mass parameters, we allow  $\alpha_{ss}$ to be different from the strong coupling constant $\alpha_s$, as in Ref.~\cite{Theu01}. Relaxing one more parameter in the Hamiltonian allows us to better reproduce the experimental results for the mass spectrum of the light baryons. All the parameters involved in the model are given in Table \ref{tab: quark core param}. The comparison with the experimental results for conventional baryons is shown in Table~\ref{tab:mass fit on light baryons}.

\begin{table}[h!]
    \centering
    \begin{tabular}{lr}
    \hline\hline
        Parameters  & \hspace{3mm} Fitted values \\
               \hline
        $m_n$ & $0.337 $\space GeV \\
        $\sigma$  & $0.1215$\space GeV$^{2}$\\
        $\alpha_{s}$ & $0.57 $\\
        $\alpha_{ss}$ & $0.239 $\\
        $\Lambda$ & $0.82$\space GeV\\
        $V_0$ & $-0.409$\space GeV\\ 
        \hline\hline
    \end{tabular}
\caption{Parameters associated with the quark core potential. These are fitted so that the eigenvalues of the Hamiltonian agree with the experimental masses of the nucleon and the $\Delta$.}
\label{tab: quark core param}
\end{table}

\begin{table}[h!]
     \centering
     \begin{tabular}{llr}
     \hline\hline
      Experimental masses \hspace{5mm}  & $J^{P}$\hspace{5mm} & Model masses\\[0.5mm]
         \hline\\[-3.5mm]
        $N_{939}$ & $\frac{1}{2}^+$ &  $941$\\
        $N_{1440}$ & $\frac{1}{2}^+$ & $1618$\\
        $N_{1710}$ & $\frac{1}{2}^+$ & $1879$\\
        $\Delta_{1232}$ & $\frac{3}{2}^+$ & $1232$\\
        $\Delta_{1600}$ & $\frac{3}{2}^+$ & $1866$ \\
        $N_{1520}-N_{1535}$ & $\frac{3}{2}^--\frac{1}{2}^-$ & $1531$\\
        $N_{1650}-N_{1675}-N_{1700}$ & $\frac{1}{2}^--\frac{5}{2}^--\frac{3}{2}^-$ & $1683$\\
        $\Delta_{1620}-\Delta_{1700}$ & $\frac{1}{2}^--\frac{3}{2}^-$ & $1648$\\
        $N_{1680}-N_{1720}$ & $\frac{5}{2}^+-\frac{3}{2}^+$ & $1883$\\[1mm]
        
        \hline \hline
     \end{tabular}
     \caption{Comparison of the light baryon spectrum with experimental masses. Model parameters are fitted to reproduce at best the $N_{939}$ and $\Delta_{1232}$. Resulting values are grouped in Table~\ref{tab: quark core param}. All masses are given in MeV.}
     \label{tab:mass fit on light baryons}
 \end{table}
 
Now that the parameters are fixed, we can diagonalize  \eqref{quark core ham} in the harmonic oscillator basis~\cite{Chev24}. To do that, we first need to evaluate the action of the color and spin operators $\bm{F}_i.\bm{F}_j$  and $\bm{s}_i.\bm{s}_j$ on the trial states, a task that requires investigating their spin and color part. This is not as easy as for potentials associated with ordinary baryons. Indeed, since we are dealing with identical fermions, the quark core must be antisymmetrised under exchange of particles. However, as indicated in~\eqref{irrep of tensor product of 3 rep}, the quark core can lie in two different color octet representations, each of which satisfies mixed exchange symmetry properties. This renders the construction of proper basis trial states much more intricate. Their explicit construction is therefore performed in~\ref{app:quark core states}. The present discussion aims at computing the action of the color operator $\bm{F}_i.\bm{F}_j$ on the two eight-dimensional irreducible subspaces defined in~\eqref{irrep of tensor product of 3 rep}. For the interaction between the $i$th and $j$th quarks, this operator reads
\begin{equation}
    \begin{split}
    \boldsymbol{F}_i.\boldsymbol{F}_j = \frac{1}{2}\Big((\boldsymbol{F}_{ij})^2-\boldsymbol{F}_i^2-\boldsymbol{F}_j^2 \Big),
    \label{color operator}
    \end{split}
\end{equation}
where $\bm{F}_i^2$ stands for the quadratic Casimir of the fundamental representation of $SU_c(3)$ and $\bm{F}_{ij}^2$ for the quadratic Casimir of the irreducible representations supported by the tensor product of the quarks $i$ and $j$. Table~\ref{tab:Casimir} lists the values of the quadratic Casimir for different irreducible representations of $SU_c(3)$.
 
\begin{table}[h!]
    \centering
    \setlength{\tabcolsep}{12pt}      
    \renewcommand{\arraystretch}{2.2} 
    \begin{tabular}{cccccc}
        \hline \hline
        Irrep of $SU(3)$ 
            & $\boldsymbol{3}$ 
            & $\boldsymbol{\bar{3}}$ 
            & $\boldsymbol{6}$ 
            & $\boldsymbol{8}$ 
            & $\boldsymbol{1}$ \\
        \hline
        $\boldsymbol{F}^2$  
            & $4/3$& $4/3$& $10/3$& $3$ 
            & $0$ \\
        \hline \hline
    \end{tabular}
    \caption{Quadratic Casimir eigenvalues for selected irreducible representations of the color group $SU_c(3)$.}
    \label{tab:Casimir}
\end{table}

Using the values in Table \ref{tab:Casimir} together with the decomposition \eqref{irrep of tensor product of 3 rep}, we can compute the action of the color operator~\eqref{color operator} on the mixed symmetric and antisymmetric color states, which gives
\begin{equation}
    \langle\boldsymbol{F}_i . \boldsymbol{F}_j \rangle_{MS} = \frac{1}{3} \hspace{3em} \text{and} \hspace{3em} \langle\boldsymbol{F}_i . \boldsymbol{F}_j \rangle_{MA} = -\frac{2}{3}.
    \label{color operator action}
\end{equation}
 The action of the color operator on the symmetrized basis trial states is then directly obtained from both~\eqref{color operator action} and the decomposition of these states in the non-symmetrized ones given in \ref{app:quark core states}.\\

The results for the lowest-lying eigenvalues associated with the Hamiltonian \eqref{quark core ham}, using the parameters of Table~\ref{tab: quark core param}, are listed in Table \ref{tab:quark core spectrum} for different $I_C(J_C^{P_C})$ quantum numbers. These are compared with eigenvalues for ordinary light baryons obtained with the same parameters. Note that for this study we have assumed no global orbital excitation of the quark core, i.e $L_C=0$. This means that, firstly, the global spin quantum number $S_C$ is equivalent to the total angular momentum, $J_C=S_C$. Secondly, the parity of the core is always positive. The analysis of Table~\ref{tab:quark core spectrum} shows that the masses of the color octet quark core are higher for all the possible quantum numbers and the mass gap increases when $J_C$ and $I_C$ increase. In that table, the ground states $E_0$ and first excited states $E_1$ are given even though only the ground states will be used in the next sections. Note also that, even for $L_C=0$, there are twice more accessible states for the quark cores compared to the associated ordinary baryons. This is due to the non-trivial symmetry of the color part of the core, which reduces the constraints on the exchange symmetry of the states, leading to more possible states (see \ref{app:quark core states} for an example).

 \begin{table}[h!]
     \centering
     \begin{tabular}{llcccc}
     \hline\hline
         & & \multicolumn{2}{c}{$E_0$} & \multicolumn{2}{c}{$E_1$} \\
        \hline 
        $J_C^{P_C}$& $I_C$\hspace{5mm} & Baryon& Quark core & Baryon& Quark core \\[0.5mm]
         \hline\\[-3.5mm]
        \multirow{2}{*}{$\frac{1}{2}^+$} & $\frac{1}{2}$ & $0.941$ & $1.481$ & $1.602$ & $2.089$ \\
                                         & $\frac{3}{2}$ &  & $1.615$ &  & $2.218$\\                                
        \multirow{2}{*}{$\frac{3}{2}^+$} & $\frac{1}{2}$ &  & $1.562$ & & $2.159$ \\
                                         & $\frac{3}{2}$ & $1.232$ & $2.228$ & $1.863$ & $2.695$\\[1mm]
        \hline \hline
     \end{tabular}
     \caption{Energy spectrum of the color-octet quark core $nnn$. $E_0$ and $E_1$ stand for masses of the ground and first excited states respectively.}
     \label{tab:quark core spectrum}
 \end{table}

Now that we have the spectrum of the quark core, let us consider each of the corresponding states as a unique effective particle whose spin, intrinsic parity, isospin and rest mass are provided by the total angular momentum $J_C$, parity $P_C$, isospin $I_C$ and associated mass eigenvalue. These effective cores are not considered as point-like, and their finite spatial extension will rule their interactions with the gluon. This spatial extension is considered as encoded in the obtained eigenstates. In the next section, we will turn our attention to the description of the interaction between the quark core and the constituent gluon.


\section{Description of the quark core-gluon interaction}
\label{sec:q c- g interaction}

In this section, we will describe the interactions between the quarks and the gluon within the core-gluon model. In this way, the quark core is considered as an effective particle with a finite spatial extension whose characteristics have been computed in the previous section. Given the hypothesis of universality of the color interactions, the core-gluon interaction will be treated similarly to the quark-quark one in the core, the only differences being the size effect of the core and the color charge of the sources. Note that the latter makes the entire system analogous to a two-gluon glueball and we will take advantages of that. In fact, the higher mass of the quark core could give the impression that the system looks like \emph{gluelumps} in constituent picture~\cite{buis07,Buis08}, but this possibility will not be developed in this study. The following sub-sections are devoted to the construction of a proper core-gluon potential which describes their interactions taking the above comments into account.


\subsubsection{The point-like core-gluon interaction}
\label{sec:g-g pot}

The first step of the construction is to write down a two-body Hamiltonian taking the octet charge of the color sources into account, and to fix all the parameters of the model properly. As mentioned in the introduction of this section, the core-gluon Hamiltonian will have the same shape as the quark-quark one in~\eqref{quark core ham}, that is
\begin{equation}
    \mathcal{H}_{Cg} = \sqrt{\bm{p}^2+m_C^2 } + \sqrt{\bm{p}^2+m_g^2} + V_{Cg}(r),
    \label{gg ham}
\end{equation}
with
\begin{equation}
    V_{Cg}(r) = \sigma_{Cg}\hspace{0.3em} r + \bm{F}_C\cdot\bm{F}_g \hspace{0.3em} \frac{\alpha_{s_{Cg} }}{r} - \bm{F}_C\cdot\bm{F}_g \hspace{0.3em} \frac{8 \alpha_{ss_{Cg}} \Lambda_{Cg}^3}{3 \sqrt{\pi}m_Cm_g} e^{-\Lambda_{Cg}^2r^2} \boldsymbol{s}_C\cdot\boldsymbol{s}_g + V_{0_{Cg}}.
    \label{gg pot}
\end{equation}
Above, $\bm{p}$ and $\bm{r}$ stands for the particle's momentum and relative position in the core-gluon CoMF, respectively. In Ref.~\cite{Cimi24}, the constituent gluon has a vanishing mass. In the present paper, a non-zero constituent mass $m_g$ for the gluon is used in both the kinetic energy and the spin-spin interaction, in the same way as for the core Hamiltonian~\eqref{quark core ham}. It is actually quite usual in phenomenological models~\cite{corn82,bern82,corn83,dono84,hou01,brau04,math08b}. It can be justified since, due to the gluon self-interaction, these particles acquire a dynamical mass. A recent study \cite{Ma25} has even suggested a proper gluon mass $m_g \approx \frac{1}{2}m_p \approx 0.450$ GeV (where $m_p$ stands for the proton mass) which could unify the results of different phenomenological models.\\

Contrary to the quark core case, the computation of the action of the color operator in~\eqref{gg pot}  on the hybrid baryon states is straightforward since they are color singlet states. Taking back relation~\eqref{color operator}, Table~\ref{tab:Casimir}, and reminding that the gluon and the core are both color octets which combine to form a global singlet, the mean value of the color operator reads
\begin{equation}
    \langle\boldsymbol{F}_C.\boldsymbol{F}_g \rangle = \frac{1}{2} (0-3-3) = -3.
\end{equation}
The spin-spin operator is not as easy to treat. Indeed, the hyperfine term in~\eqref{gg pot} assumes the core and the gluon to carry spin degrees of freedom. However, even massive, the gluon is shown to still have helicity degrees of freedom~\cite{chev25.1,math08}. This issue is overcome in~\ref{app:HB states}. Note also that the spin operator associated with the core, denoted $\bm{s}_C$, could be either represented by the total spin $S_C$ or the total angular momentum $J_C$ of the core, as the core is seen as a unique particle at rest. In this study, the question does not arise since $S_C=J_C$.  \\

We chose to use a universal shape for the color interaction. However, due to the simplicity of our Hamiltonian model, we observed that it is not possible to reproduce both the baryon and glueball spectra with the same set of parameters for the potential. Therefore, parameters in~\eqref{gg ham} are adapted taking advantage of the analogy with the gluon-gluon interaction. Indeed, the Hamiltonian~\eqref{gg ham} is very similar to the one used to describe two-gluon glueballs within the helicity formalism in Ref. \cite{math08,chev25.1}, with results in agreement with LQCD computations. The differences are the addition of the hyperfine term and the non-zero masses of the involved particles. The values of the string tension and the shift constant of~\cite{math08} will then be taken as references for the fit. The strong coupling parameter $\alpha_{s_{Cg}}$, as well as the new parameters $\alpha_{ss_{Cg}}$ and $\Lambda_{Cg}$ of the hyperfine term are then fitted to reproduce as much as possible the glueball spectrum obtained using LQCD in Ref. \cite{meye05,chen06}. After the fit, the obtained glueball spectrum lies within the Lattice error bars (see Table~\ref{tab:mass fit on two-gluon glueballs}). Table \ref{tab: gg param} regroups all the parameters of the model. Note that, contrary to the fitting of the parameters done in Section \ref{sec:quark core spectrum}, $\alpha_{s_{Cg}}$ and $\alpha_{ss_{Cg}}$ are not needed to be different in order to improve the fit. For the phenomenology of two-gluon glueballs, the shifting constant $V_0$ is not necessary. The potential~\eqref{gg pot} finally reads
\begin{equation}
    V_{Cg}(r) = \sigma_{Cg}\hspace{0.3em} r -3 \hspace{0.3em} \frac{\alpha_{s_{Cg} }}{r} + \frac{8 \alpha_{ss_{Cg}} \Lambda_{Cg}^3}{ \sqrt{\pi}m_Cm_g} e^{-\Lambda_{Cg}^2r^2} \boldsymbol{s}_C\cdot\boldsymbol{s}_g.
    \label{c-g pot}
\end{equation}
\begin{table}[h!]
     \centering
     \begin{tabular}{lcrrr}
     \hline\hline
      State \hspace{5mm}  & \hspace{0.2cm}Present model\hspace{0.2cm} & \hspace{2cm} LQCD~\cite{chen06} & \hspace{2cm} LQCD~\cite{meye05}\\[0.5mm]
         \hline\\[-3.5mm]
        $\ket{\Psi_1;S_+,0^+}$ & $1.596$ & $1.710\pm 0.050\pm0.080$ & $1.475\pm0.030\pm0.065$ \\
        $\ket{\Psi_2;S_+,0^+}$ & $2.766$ & $2.670\pm0.180\pm0.130$ & $2.755\pm0.070\pm0.120$\\
        $\ket{\Psi_1;S_-,0^-}$ & $2.309$ & $2.560\pm0.035\pm 0.120$ & $2.250\pm0.060\pm0.100$ \\
        $\ket{\Psi_2;S_-,0^-}$ & $3.141$ & $3.640\pm0.060\pm0.180$ & $3.370\pm0.150\pm0.150$ \\
        $\ket{\Psi_1;D_+,2^+}$ & $2.366$ & $2.390\pm0.030\pm0.120$ & $2.150\pm0.030\pm0.100$ \\
        $\ket{\Psi_1;S_+,2^+}$ & $3.071$ & $...$ & $2.880\pm0.100\pm0.130$ \\
        $\ket{\Psi_1;S_-,2^-}$ & $3.074$ & $3.040\pm0.040\pm0.150$ & $2.780\pm0.050\pm0.130$ \\
        $\ket{\Psi_2;S_-,2^-}$ & $3.790$ & $3.890\pm0.040\pm0.190$ & $3.480\pm0.140\pm0.160$ \\
        $\ket{\Psi_1;D_-,3^+}$ & $3.367$ & $3.670\pm0.050 \pm0.180$ & $3.385\pm0.090\pm0.150$ \\
        \hline \hline
     \end{tabular}
     \caption{Comparison of the two-gluon glueball spectrum computed using the the LMM with an Hamiltonian structure as~\eqref{gg ham} and the parameters in Table~\ref{tab: gg param} with the LQCD results of~\cite{chen06} and~\cite{meye05}. Notations for the states are taken from~\cite{chev25.1}. Energies are provided in GeV}.
     \label{tab:mass fit on two-gluon glueballs}
 \end{table}
\begin{table}[h!]
    \centering
    \begin{tabular}{lr}
    \hline\hline
        Parameters  & \hspace{3mm} Fitted values \\
               \hline 
        $m_g$&$0.450$ GeV\\
        $\sigma_{Cg}$  & $0.416\text{ GeV}^{2}$\\
        $\alpha_{s_{Cg}}$ & $0.492 $\\
        $\alpha_{ss_{Cg}}$ & $0.492 $\\
        $\Lambda_{Cg}$ & $0.162$ GeV \\
        $V_{0_{Cg}}$ & $0 \text{ GeV}$\\ [0.5mm]
        \hline\hline
    \end{tabular}
    \caption{Parameters associated with the core-gluon Hamiltonian \eqref{gg ham}. These are chosen so that the two-gluon glueball mass spectrum fit the Lattice results of \cite{meye05,chen06}. }
    \label{tab: gg param}
\end{table}


\subsubsection{Size effect of the core}
\label{sec:size effect}

\begin{figure}
    \centering
    \includegraphics[width=0.5\linewidth]{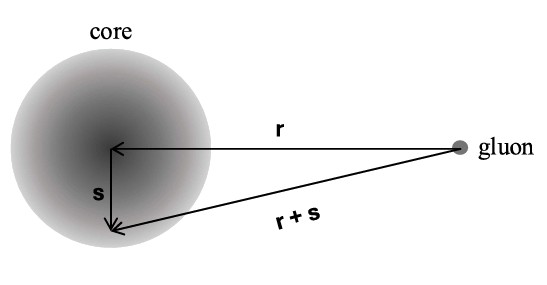}
    \caption{Schematization of the quark core-gluon convoluted interaction, adapted from \cite{tour25}.}
    \label{fig:core-gluon}
\end{figure}

 As discussed in Section~\ref{sec:quark core spectrum}, the quark core has a complex internal structure, which gives a spatial extent to its color charge distribution. This size is expected to modify the core-gluon interaction. Indeed, unlike the potential~\eqref{c-g pot}, which depends on the norm of the relative position between the core and the gluon, the true interaction should link the gluon with all the internal quarks.  This issue first appeared in the study of ordinary baryons in the quark-diquark picture. In Ref.~\cite{tour25}, the authors suggest to overcome it by replacing the  quark-diquark potential by its convolution with the color density of the diquark. This led to results with very good agreements with a full three-body description, for both light and heavy sectors. The same method will then be adopted here. Mathematically, the convolution reads
\begin{equation}
    \tilde{V}_{Cg}(r) = \int d^3s \frac{\rho(\boldsymbol{s})}{3} V_{Cg}(|\boldsymbol{r}+\boldsymbol{s}|),
    \label{conv pot}
\end{equation}
where $V_{Cg}$ is given in~\eqref{c-g pot} and $\rho/3$ is the normalized color density of the core. The integration is performed with respect to the position relative to the CoMF of the core. Figure \ref{fig:core-gluon} depicts a schematic visualization of the core-gluon interaction defined by~\ref{conv pot}.
In order to determine the convoluted potential, we first need to compute the color density of the core. This density is actually equal to the quark one because all quarks inside the core possess the same color charge. For three-body cores, the quark density function is defined as follows
\begin{equation}
    \rho(\boldsymbol{q}) = \sum_{j=1}^3 \iiint \psi^*(\boldsymbol{q}_1,\boldsymbol{q}_2,\boldsymbol{q}_3)\delta^3(\boldsymbol{q}-\boldsymbol{q}_j)\psi(\boldsymbol{q}_1,\boldsymbol{q}_2,\boldsymbol{q}_3) d^3q_1d^3q_2d^3q_3,
\end{equation}
where $\psi(\bm{q}_1,\bm{q}_2,\bm{q}_3)$ is the normalized wave function of the core in some inertial reference frame. But we want $\rho$ to be a function of the position relative to the CoMF, $\boldsymbol{s}$. Thus, we will define $\boldsymbol{s} = \boldsymbol{q}-\boldsymbol{R} $ and $\boldsymbol{s}_j = \boldsymbol{q}_j - \boldsymbol{R}$, where $\bm{R} = \frac{\bm{q}_1+\bm{q}_2+\bm{q}_3}{3}$ is the position of the center of mass of the three identical quarks in the inertial reference frame. In this way, by changing the $\bm{s}_j$ coordinates to the so-called dimensionless Jacobi coordinates $\{\boldsymbol{x},\boldsymbol{y}\}$ (defined in~\ref{app:quark core states}), the quark density is expressed as an integral over the internal part of the core wave function, $\psi_{int}(\bm{x},\bm{y})$,
\begin{equation}
    \rho(\boldsymbol{s}) = \sum_{j=1}^3 \iint \psi_{int}^*(\boldsymbol{x},\boldsymbol{y})\delta^3(\boldsymbol{s}-\boldsymbol{s}_j(\boldsymbol{x},\boldsymbol{y})) \psi_{int}(\boldsymbol{x},\boldsymbol{y})d^3x d^3y,
    \label{rho}
\end{equation}
where the integral over the center of mass coordinate has been dropped. The explicit expression of $\psi_{int}$ is given by the resolution of the core dynamics with OBE, and is described in~\ref{app:quark core states}. Inserting it in \eqref{rho} and performing the two integrals lead to
\begin{equation}
\begin{split}
    \rho_{LM}(s,\theta_s,\phi_s) &= \frac{9\sqrt{3}}{a^3} \sum_{n_x,l_x,m_x}\sum_{n_y,l_y,m_y}\sum_{n'_y,l'_y,m'_y}
    C_{n_x l_x n'_yl'_y}^* C_{n_xl_xn_yl_y} ( l_xm_xl'_ym'_y|LM) ( l_x m_x l_y m_y | LM )\\
    &\times \mathcal{R}_{n'_yl'_y}(\frac{\sqrt{3}}{a}s)\mathcal{R}_{n_yl_y}(\frac{\sqrt{3}}{a}s)Y_{l'_ym'_y}^{*}(\pi-\theta_s,\pi+\phi_s)Y_{l_ym_y}(\pi-\theta_s,\pi+\phi_s),
    \end{split}
    \label{quark density}
\end{equation}
In this expression, the functions $\mathcal{R}_{nl}$ and $Y_{lm}$ are the radial solutions of the $3D$ harmonic oscillator~\cite{Chev24} and the spherical harmonics, respectively, the factors $C_{n_xl_xn_yl_y}$ are expansion coefficients and $a$ is a variational scale parameter. The values of the latters are provided by OBE. Note that the construction of the quark density presented above implicitly assumes a non-relativistic treatment of the change of reference frame. The core wave function, and hence the resulting density $\rho(\bm{s})$, is computed in the CoMF of the core itself, whereas the convolution~\eqref{conv pot} combines it with the core-gluon potential, naturally defined in the CoMF of the core-gluon system. A fully covariant construction would require boosting the internal wave function of the core into the latter frame before performing the convolution, which considerably complicates the calculation. This approximation, reminiscent of the Bakamjian-Thomas construction of relativistic dynamics~\cite{Keis91}, is justified by the large mass of the core: its recoil velocity in the core-gluon CoMF remains small compared to the internal motion of the quarks, so relativistic corrections to the density are expected to be subdominant.

\vspace{1em}

\begin{figure}[h!]
    \centering
    \includegraphics[width=1\linewidth]{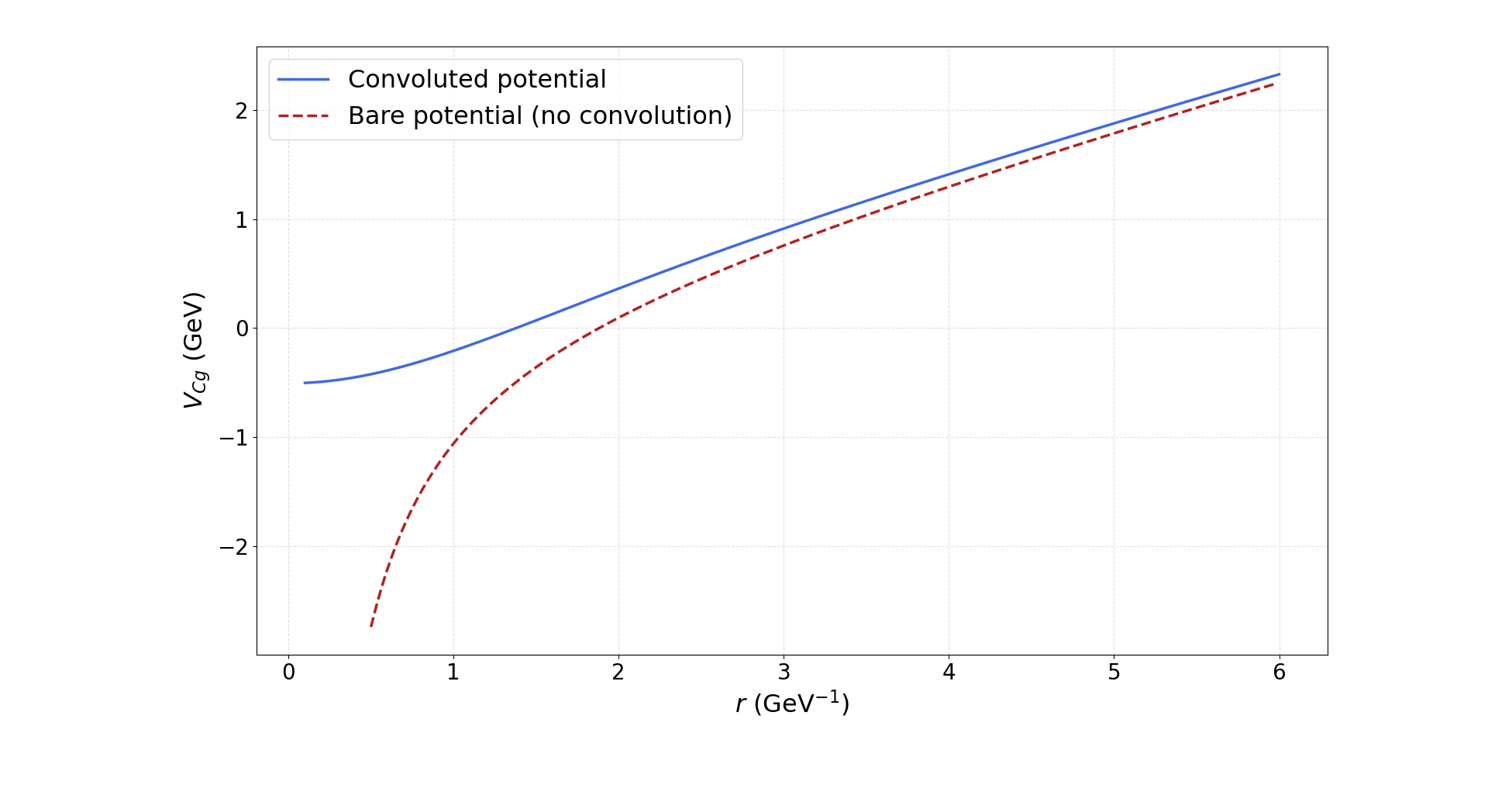}
    \caption{Comparison of the convoluted and non-convoluted Cornell part of the core-gluon potential \eqref{c-g pot} for $L_C=0, I_C=1/2, J_C=S_C=1/2$. See text for the details.}
    \label{fig:conv vs non-conv}
\end{figure}

Now that we have an expression for the quark density, we can plug it into relation \eqref{conv pot} to find the convoluted potential.  
Mimicking the method of computation for the convoluted potential in~\cite{tour25}, the relation~\eqref{quark density} has to be slightly modified in order to cancel the dependence on the magnetic quantum number of the core, present for $L\neq 0$. It is done by averaging over all possible values of $M$. Even though only cores with $L=0$ are considered in this work, we give the general formula,
\begin{equation}
    \tilde{V}_{{Cg}}^L(r) = \frac{1}{(2L+1)} \sum_{M=-L}^L \int d^3s \frac{\rho_{LM}(\boldsymbol{s)}}{3} V_{Cg}(|\boldsymbol{r}+\boldsymbol{s}|).
    \label{conv pot of L}
\end{equation}
Note that the dependence of $\tilde V_{{Cg}}^L$ on the orbital quantum number of the core is expected, given that this quantum number is strongly correlated with the size of the system. Now, inserting~\eqref{quark density} in \eqref{conv pot of L} and performing the integrals in spherical coordinates, we obtain
\begin{equation}
    \tilde{V}_{{Cg}}^L(r) = \frac{3\sqrt{3}}{2a^3 } \sum_{n_x,l_x,n_y,l_y,n'_y} C_{n_xl_xn'_yl_y } C_{n_xl_xn_yl_y} \int ds   s^2 \mathcal{R}_{n'_yl_y}(\frac{\sqrt{3}}{a}s) \mathcal{R}_{n_yl_y}(\frac{\sqrt{3}}{a}s) I(r,s),
    \label{core-gluon conv pot le final}
\end{equation}
where
\begin{equation}
    I(r,s) = \int_{-1}^{1}d\mu V_{Cg}(\sqrt{r^2+s^2+2rs\mu}).
\end{equation}
The simplification of the Clebsh-Gordan coefficients and the spherical harmonics, leading to~\eqref{core-gluon conv pot le final}, is done using the following properties~\cite{vars88}:
\begin{equation}
\begin{split}
    \sum_{m_x,M}(l_x m_xl_ym_y|LM)(l_xm_xl_y'm_y'|LM)& = \frac{2L+1}{2 l_y +1}\delta_{l_y'l_y}\delta_{m_y'm_y}\\
    \sum_{m}|\mathcal{Y}_{lm}(\theta_s,\phi_s)|^2=\frac{2l+1}{4\pi}.
\end{split}
\end{equation}
Note that the function $V_{Cg}$ in~\eqref{core-gluon conv pot le final} is the non-convoluted core-gluon potential, whose expression is given by~\eqref{c-g pot}. It can be interesting to visualize the effect of the convolution on the potential shape between the quark core and the gluon. Figure \ref{fig:conv vs non-conv} compares the convoluted potential $\tilde{V}_{{Cg}}^L$ with its non-convoluted version, given by \eqref{c-g pot}, omitting the hyperfine term, for which the action on the hybrid baryon states is not direct, for some specific quantum numbers $L_C=0, I_C=1/2$ and $J_C = S_C=1/2$. We can see that the convolution tends to regularize the potential at short range, as expected since at this range the potential enters the sector where the quark density is large. At long range, the two potentials merge, as expected since the quark-core becomes more and more point-like. Due to the linear long range part of the non-convoluted potential, this merging is slow. \\


\section{Solving the core-gluon dynamics}
\label{sec:HB spectrum}

In this section, the core-gluon dynamic will be solved numerically. To do that, we first have to build the two-body states representing the hybrid baryons. Due to the presence of the gluon in the system, even though it has a constituent mass, this construction has to be realized within the helicity formalism~\cite{Jaco59}. Indeed, in Ref.~\cite{math08}, it is shown that the use of helicity is necessary to reproduce the LQCD glueball spectrum. Then, we will use a version of the LMM~\cite{sema01,lacr12} adapted to handle helicity states~\cite{Chev25.2} to obtain the associated mass spectrum.


\subsection{Hybrid baryon states within the helicity formalism}
\label{sec: helicity formalism}

In the helicity formalism, particle states are characterized by their helicity~\cite{Jaco59}, which is a natural quantum number for massless particles~\cite{Wein95}. In this work, as we want to predict the mass spectrum of hybrid baryons with quantum numbers $J^P$, the construction of such states should be done in the CoMF. In what follows, we will only focus on the spatial and spin part of the physical states, since the color part has already been treated independently in Section~\ref{sec:g-g pot}. Ref.~\cite{Jaco59,chev25.1} provide such a construction, leading to a complete set of two-body helicity states with definite total angular momentum $J,\,M$ and helicities $\lambda_C,\,\lambda_g$
\begin{equation}
    |p,JM,\lambda_C\lambda_g;J_C\rangle = \sqrt{\frac{2J+1}{4\pi}} \int \diff\cos{\theta} \diff\phi D_{M \lambda_C-\lambda_g}^{J*}(\phi,\theta,0)|p\theta\phi,\lambda_C\lambda_g;J_C\rangle.
    \label{J-helicity states}
\end{equation}
Above, the states $|p\theta\phi,\lambda_C\lambda_g;J_C\rangle$ are defined as eigenstates of the relative momentum, core total spin, core helicity and gluon helicity operators. The associated eigenvalues are given by the quantum numbers  $p, \theta, \phi,J_C, \lambda_C$ and $\lambda_g$, respectively (the relative momentum is given in spherical coordinates). Note that for a massive core and a massless gluon $\lambda_C\in \llbracket -J_C,J_C\rrbracket$ and $\lambda_g\in\{-1,1\}$. Let us remind that for this study, $J_C = S_C$ as $L_C=0$. The states defined in~\eqref{J-helicity states} are by construction eigenstates of the total angular momentum operators with corresponding eigenvalues given by $J,\,M$. They have the following normalization
\begin{equation}
    \bra{p',J'M',\lambda'_C\lambda'_g;J_C} p,JM,\lambda_C\lambda_g;J_C\rangle = \frac{4 w_C(p)w_g(p)}{p^2}\delta(p'-p)\delta_{J'J}\delta_{M'M}\delta_{\lambda'_C\lambda_C}\delta_{\lambda'_g\lambda_g},
\end{equation}
where $w_C(p) = \sqrt{p^2+m_C^2}$ and $w_g(p) = \sqrt{p^2+m_g^2}$ are the energies of the core and the gluon in the CoMF, respectively. The function $D_{M  \lambda_c-\lambda_g}^{J*}$ in~\eqref{J-helicity states} is a Wigner $D$ matrix~\cite{vars88}. Note that the presence of such a Wigner $D$ matrix constraints the possible values for $J$ so that $J \geq |\lambda_C-\lambda_g|$. Although the states~\eqref{J-helicity states} carry a definite total angular momentum, they do not possess a well-defined parity. Indeed, the action of the parity operator $\Pi$ on these states reads~\cite{Jaco59}
\begin{equation}
\Pi |p,JM,\lambda_C\lambda_g;J_C\rangle = \eta_C \eta_g (-1)^{J-J_C-1} |p,JM,-\lambda_C-\lambda_g; J_C\rangle,
\end{equation}
where $\eta_C$ and $\eta_g$ denote the intrinsic parities of the core and the gluon, respectively. Note that $\eta_C$ is here chosen as the parity quantum number defined in Sec.~\ref{sec:quark core spectrum}. It turns out, however, that suitable linear combinations of these states do possess definite parity. Specifically, one can define
\begin{equation}
|p,J^{P_i},\lambda_C\lambda_g;J_C\rangle := \frac{1}{\sqrt{2}}\left(|p,JM,\lambda_C\lambda_g;J_C\rangle + (-1)^{i}|p,JM,-\lambda_C-\lambda_g;J_C\rangle\right),
\label{physical states}
\end{equation}
with $i=0,1$. The corresponding parity eigenvalues are then given by
\begin{equation}
P_i = (-1)^{i}\eta_C\eta_g (-1)^{J-J_C-1}.
\end{equation}
The two-body states defined in \eqref{physical states} will be our basis states of the spatial and spin part of the Hilbert space carrying the physical hybrid baryon states. Following Ref.~\cite{chev25.1}, we will define a generic hybrid baryon state with definite $J^P$ quantum numbers as
\begin{equation}
    |\Psi,J^{P_i},\lambda_C\lambda_g;J_C\rangle = \int \frac{p^2dp}{4w_C(p)w_g(p)}\Psi(p)|p,J^{P_i},\lambda_C\lambda_g;J_C\rangle,
    \label{HB states}
\end{equation}
where the function $\Psi(p)$ should be interpreted as a radial helicity-momentum wave function, having the following normalization condition so that $\langle\Psi,J^{P_i},\lambda_C\lambda_g;J_C|\Psi,J^{P_i},\lambda_C\lambda_g;J_C\rangle=1$~\cite{chev25.1}:
\begin{equation}
    \int \frac{p^2 \diff p}{4 w_1(p) w_2(p)} |\Psi(p)|^2 = 1.
\end{equation}


\subsection{The Lagrange Mesh Method to solve the hybrid baryon Hamiltonian}
\label{sec: LMM for HB}

The LMM is a numerical method to solve eigenvalue problems. It is based on a truncated basis expansion with definite trial states, whose matrix elements are computed by mean of Gauss-Laguerre quadratures \cite{sema01}. This method was initially developed for problems in position space representation. The physical states of hybrid baryons~\eqref{HB states} being constructed in momentum space representation, we should use an adapted LMM for the present problem~\cite{lacr12}.  However, the methodology employed in~\cite{lacr12} is not suited for the Cornell potential. This LMM was recently extended in \cite{Chev25.2} for such a potential and tested on two-gluon glueballs. Let us shortly review the main calculation steps, emphasizing on the characteristics of the present system.

\vspace{1em}

In the LMM adapted to our problem, we expand the generic physical states given in \eqref{HB states} as a linear combination of trial states:
\begin{equation}
    |\Psi,J^{P_i},\lambda_g;J_C\rangle = \sum_{j=1}^N\sum_{\lambda_C=1/2}^{J_C} C_{j\lambda_C}|f_j,J^{P_i},\lambda_C\lambda_g;J_C\rangle,
    \label{LM decomposition}
\end{equation}
where $C_{j\lambda_C}$ are the expansion coefficients to determine, $N$ is the number of mesh points, which is the same for each $\lambda_C$. The sum on $\lambda_C$ is constrained by the condition $J\geq |\lambda'_C-\lambda_g|$ and should not cover negative values because of~\eqref{physical states}. The trial states in~\eqref{LM decomposition} are defined by 
\begin{equation}
    |f_j,J^{P_i},\lambda_C\lambda_g;J_C\rangle = \int\frac{p^2 dp}{2 \sqrt{w_C(p)w_g(p)}}\frac{f_j(p/h)}{p\sqrt{h}}|p,J^{P_i},\lambda_C\lambda_g;J_C\rangle,
    \label{HB trial states}
\end{equation}
where the functions $f_j$ are regularized Lagrange functions defined in \cite{Chev25.2}, and $h$ is a size parameter. The eigenvalue problem then becomes 
\begin{equation}
    \sum_{k,\lambda'_C}C_{k\lambda'_C} \langle f_j,J^{P_i},\lambda_C\lambda_g;J_C|\mathcal{H}_{HB}|f_k,J^{P_i},\lambda'_C\lambda_g;J_C\rangle = E \space  C_{j\lambda_C}.
    \label{eigenvalue problem}
\end{equation}
To solve this system, we first need to evaluate the Hamiltonian matrix of order $(N \times N_{\lambda_C})^2$, with $N_{\lambda_C}$ the number of possible helicities for the quark core. This computation is described in \ref{app:HB states}. Note that the helicity of the quark core not being fixed, channels with different $\lambda_C$ could be coupled with non-zero mixing coefficients $C_{j\lambda_C}$. Then, by diagonalizing the resulting Hamiltonian matrix, we find the eigenenergies $E$ of the hybrid baryon that we want to study, and the eigencoefficients of the expansion $C_{j\lambda_C}$, which allows us to get the hybrid baryon wave function.

\section{Light hybrid baryons spectrum}
\label{sec: spectrum}

Let us now bring together all the elements that we have discussed until now, and write down explicitly the complete hybrid baryon Hamiltonian of our model. It is given by
\begin{equation}
    \mathcal{H} _{HB} = \sqrt{\boldsymbol{p}^2+m_C^2} + \sqrt{\boldsymbol{p}^2+m_g^2} + \tilde{V}_{Cg}^L(r),
    \label{HB hamiltonian}
\end{equation}
where $\tilde{V}_{Cg}^L(r)$ is the convoluted potential given by \eqref{core-gluon conv pot le final}. This last function has to be evaluated numerically, and since we use a mesh method to solve the Hamiltonian the evaluation is done only at the mesh points. The mass parameter $m_C$ is taken as the mass of the quark core, computed in Section \ref{sec:quark core spectrum}. All the other parameters are provided in Table~\ref{tab: gg param}. The results are presented in Table \ref{tab:HB spectrum} for $J_C = 1/2, 3/2$, $J^P = 1/2^{\pm},3/2^{\pm}$ and for the $N$-like and the $\Delta$-like cores, corresponding to the total isospin $I_C=1/2$ and $I_C = 3/2$, respectively.

\setlength{\tabcolsep}{10pt} 

\begin{table}[h!]
\centering
\begin{tabular}{cccccc}
\hline\hline
$J_C$ & $I_C$ & $m_C$ & $J^P$ & $M_0$ & $M_1$ \\
\hline
\multirow{8}{*}[-1.7em]{$\frac{1}{2}$} 
& \multirow{4}{*}{$\frac{1}{2}$} 
& \multirow{4}{*}{$1.481$} 
& $\frac{1}{2}^-$ & $3.293$ & $4.138$ \\
& & & $\frac{1}{2}^+$ & $3.444$ & $4.219$ \\
& & & $\frac{3}{2}^-$ & $3.302$ & $3.954$ \\
& & & $\frac{3}{2}^+$ & $3.452$ & $3.864$ \\
\\[-4pt]
\cline{2-6}
\\[-4pt]
& \multirow{4}{*}{$\frac{3}{2}$}
& \multirow{4}{*}{$1.615$} 
& $\frac{1}{2}^-$ & $3.411$ & $4.250$ \\
& & & $\frac{1}{2}^+$ & $3.561$ & $4.330$ \\
& & & $\frac{3}{2}^-$ & $3.420$ & $4.066$ \\
& & & $\frac{3}{2}^+$ & $3.568$ & $3.977$ \\[1mm]
\hline
\multirow{8}{*}[-1.7em]{$\frac{3}{2}$} 
& \multirow{4}{*}{$\frac{1}{2}$} 
& \multirow{4}{*}{$1.562$} 
& $\frac{1}{2}^-$ & $ 3.590$ & $ 4.162$ \\
& & & $\frac{1}{2}^+$ & $3.699$ & $ 4.094$ \\
& & & $\frac{3}{2}^-$ & $3.593$ &  $4.163$ \\
& & & $\frac{3}{2}^+$ & $3.701$& $4.094$ \\
\\[-4pt]
\cline{2-6}
\\[-4pt]
& \multirow{4}{*}{$\frac{3}{2}$}
& \multirow{4}{*}{$2.228$} 
& $\frac{1}{2}^-$ & $4.202$ &  $4.748$ \\
& & & $\frac{1}{2}^+$ & $4.304$ & $4.684$ \\
& & & $\frac{3}{2}^-$ & $4.204$ & $4.749$ \\
& & & $\frac{3}{2}^+$ & $4.306$ & $4.685$ \\
\\[-8pt]
\hline\hline
\end{tabular}
\caption{Mass spectrum of light hybrid baryons. $J_C$ and $I_C$ stand for the spin and isospin quantum numbers of the core and $m_C$ stands for its mass, computed in Section~\ref{sec:quark core spectrum}. $J^P$ stands for the quantum numbers of the hybrid baryons, and $M_0$ and $M_1$ for the mass of their associated ground and first excited states, respectively. Masses are given in GeV. }
\label{tab:HB spectrum}
\end{table}
We see that negative parity states are predicted to be lighter than their positive parity homologue. This means that the lightest hybrid baryon would have negative parity. Note also that the mass splitting between the hybrid $N$ and $\Delta$ is about $120\text{ MeV}$, which is two times less than for the conventional baryons. Then, even for a quark core with no orbital excitation, the hybrid baryon spectrum already looks really rich, with several states with the same radial excitation bringing the same physical quantum numbers $I(J^P)$.  This suggests that the physical hybrid baryon states could be in a quantum superposition of all these states with the same quantum numbers. Accessing to the associated mixing coefficients would require a deeper study. Finally, we have remarked that the hyperfine term in the core-gluon potential does not contribute significantly to the energy of the hybrid baryons. \\

In order to perform a deeper analysis of the results, let us compare them with those obtained using other theoretical methods, such as LQCD or QCD sum rules. From the Lattice side, Ref.~\cite{Dudek12} focused on the positive parity hybrid spectrum for the $N$ and the $\Delta$ (see Fig. $1$ of \cite{Dudek12}), since negative parity states turned out significantly higher in mass. For the $N$, hybrid states were found to have a mass between $2.5 \text{ GeV}$ and $3\text{ GeV}$ for the states with $J^{P}=1/2^+,3/2^+$ and $5/2^+$, which is about $0.9\text{ GeV}$ lower than our results. For the delta, hybrid states were found to have a mass between $2.8\text{ GeV}$ and $3\text{ GeV}$ for the states with $J^P=1/2^+$ and $J^P=3/2^+$, which is about $0.8\text{ GeV}$ lower than our results. We can thus notice that, though a similar energy gap between states with different quantum numbers, the overall mass range of the spectrum obtained in LQCD is not in agreement with the one obtained in this study.

\begin{figure}[h!]
    \centering

    \begin{subfigure}[t]{0.85\textwidth}
        \centering
        \includegraphics[width=\textwidth]{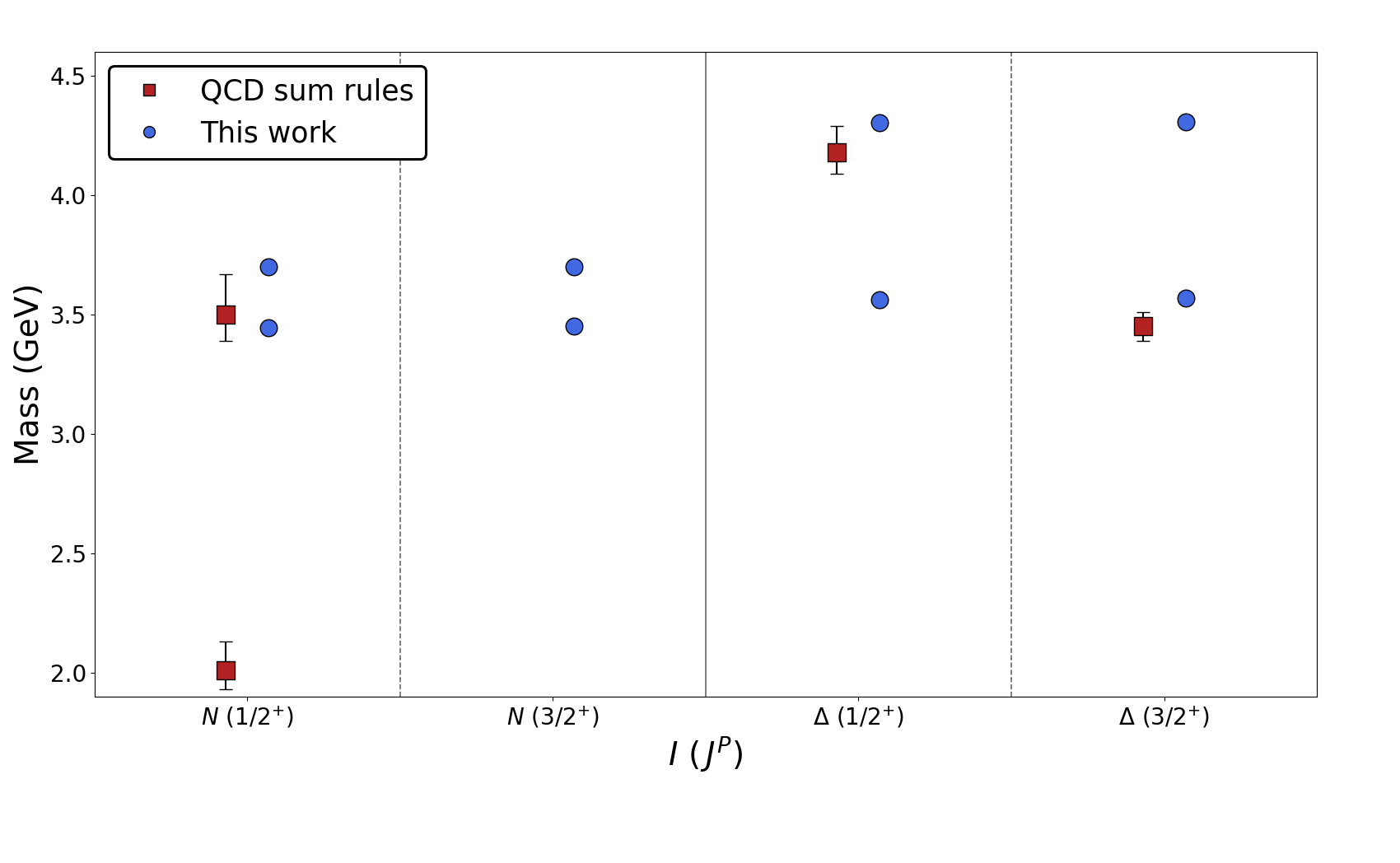}
        \caption{Positive parity}
        \label{fig:positive parity}
    \end{subfigure}
    \hfill
    
    \begin{subfigure}[t]{0.85\textwidth}
        \centering
        \includegraphics[width=\textwidth]{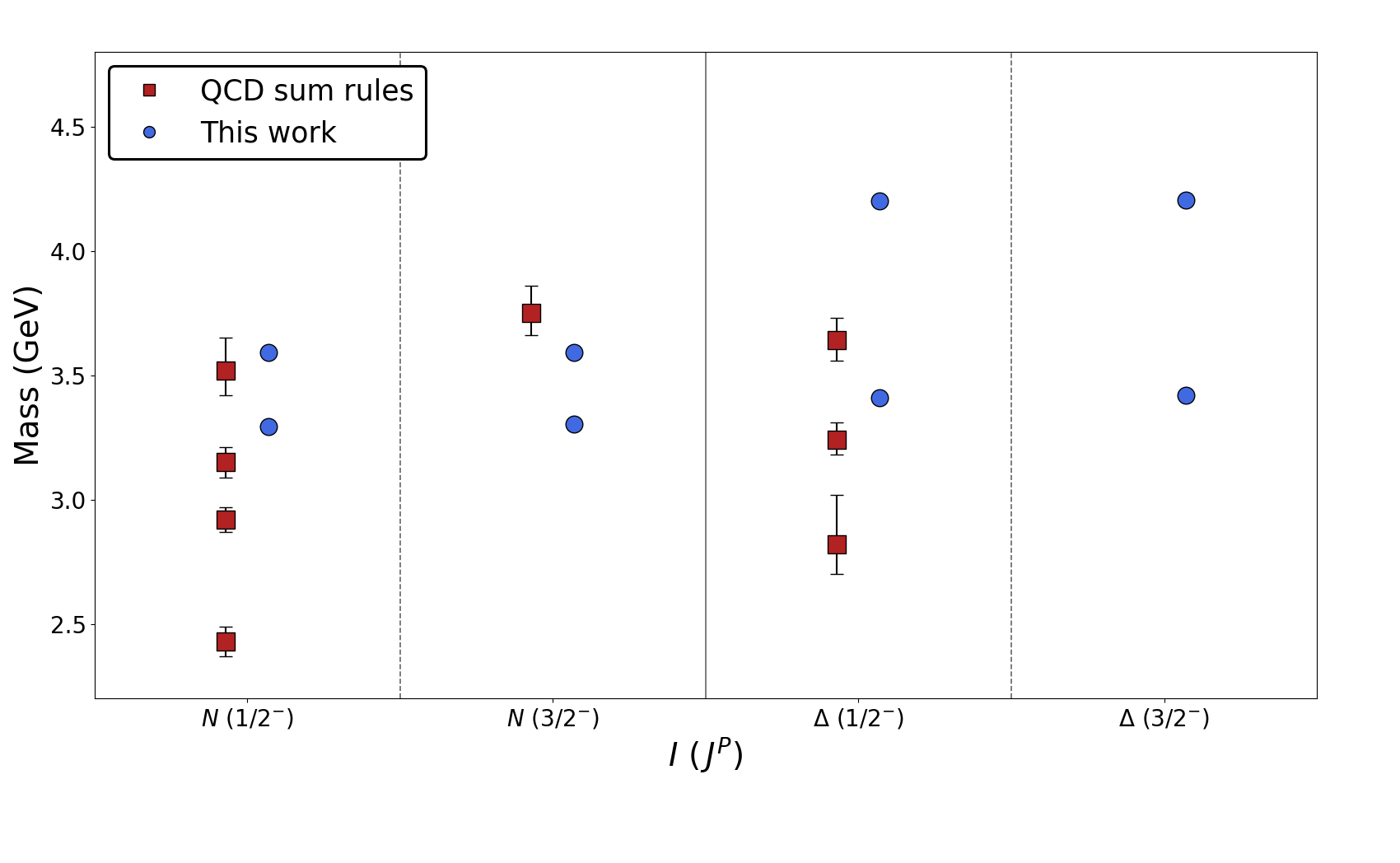}
        \caption{Negative parity}
        \label{fig:negative parity}
    \end{subfigure}

    \caption{
        Mass spectrum of light hybrid baryons.
       The blue circles correspond to predictions of the core-gluon model while the red squares correspond to those of the QCD sum rule method, taken from \cite{Wang26}.
       The different values of energy for same quantum numbers come from distinct reasons in the two models. For the core-gluon model, it comes from different internal structures, while for the QCD sum rule method, it comes from the different interpolating currents.
        }
    \label{fig:core-gluon vs QCD sum rule}
\end{figure}
From the sum rule side now, both positive and negative parity hybrid spectra are determined for $N$ and $\Delta$ (see Fig. $3$ of \cite{Wang26}). Figure \ref{fig:core-gluon vs QCD sum rule} shows the comparison between these spectra and the spectrum from Table~\ref{tab:HB spectrum} obtained in this study. We see that for the $\Delta_{J^P}=\Delta_{1/2^+}$ and $\Delta_{3/2^+}$,  which turn out to be the channels with the most stable QCD sum rules, the masses given by both methods differ by less than $100$ MeV. However, for the rest of the channels, the agreement is less obvious. Indeed, the energy range spanned by the different currents is wider for the channels $N_{1/2^+},N_{1/2^-}$ and $\Delta_{1/2^-}$, which render all comparison difficult. What we can say is that the energy ranges predicted by our method at least overlap those predicted by the QCD sum rule method. Besides, it is important to note that, contrary to our results, those of Ref.~\cite{Wang26} predict the lightest hybrid baryon to have a positive parity, which was also the case with the Lattice results. It turns out that this disagreement was also pointed out in the study of gluelumps within the quark model \cite{Buis08}. In fact, it is not unreasonable since, as mentioned in Section~\ref{sec:q c- g interaction}, the constituent approach of gluelumps is very similar to the core-gluon model for hybrid baryons. Indeed, in Ref.~\cite{Buis08}, the author considers the gluelump as a system composed of a single transverse constituent gluon interacting with a static color-octet source via a QCD-based potential like \eqref{gg pot}. This is actually the limit of the core-gluon model for a quark core with infinite mass and $0^{+}$ quantum numbers. It is not the case for our hybrid baryons but the core having a mass more than three time heavier than the constituent gluon, the system can be seen as approaching the one for gluelumps. In Ref.~\cite{Buis08}, the author suggested a way to overcome this disagreement by introducing a phenomenological parity splitting mass term in the potential which is justified as arising from instanton-induced interactions. This has not been done in the present work, but it could be interesting to keep in mind for future improvement of our model. Another way to overcome this parity disagreement could be to invert the intrinsic parity of the constituent gluon. Although we did not found any theoretical mechanism justifying the use of a positive parity gluon in a constituent model, other studies use a gluon field with $J^P=1^+$~\cite{Dono81,Chow99}. This modification would exchange the positive and negative-parity spectrum. It would lead to a better agreement with the LQCD and QCD sum rules results, but, given that the hybrid baryon spectrum presented in Table~\ref{tab:HB spectrum} is almost parity degenerated, our results and conclusions would not fundamentally change in practice.

\subsubsection{Extension of the model to heavy hybrid baryons}

\begin{figure}[h!]
    \centering

    \begin{subfigure}[t]{0.85\textwidth}
        \centering
        \includegraphics[width=\textwidth]{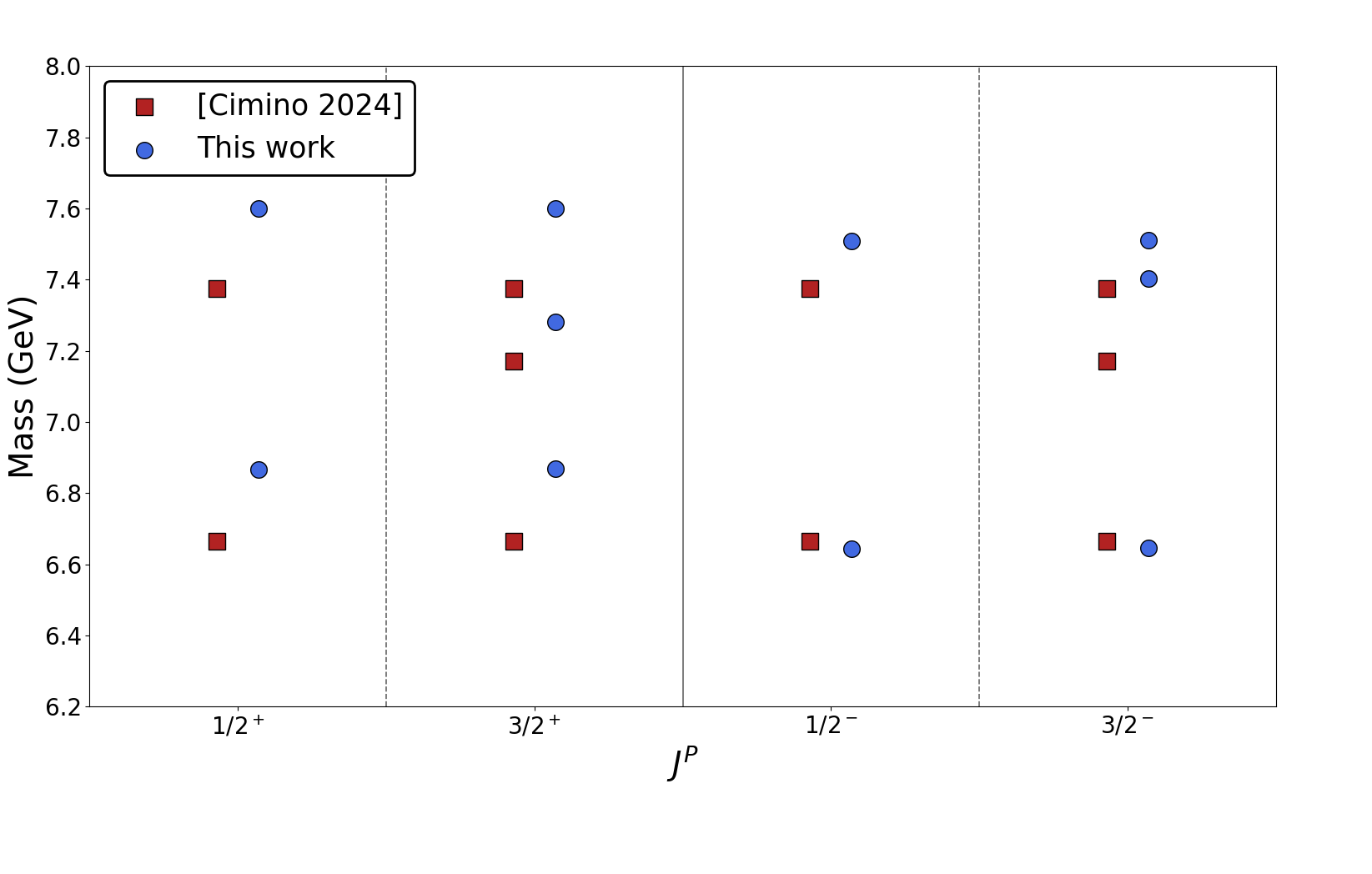}
        \caption{$cccg$ mass spectrum}
        \label{fig:cccg}
    \end{subfigure}
    \hfill
    \begin{subfigure}[t]{0.85\textwidth}
        \centering
        \includegraphics[width=\textwidth]{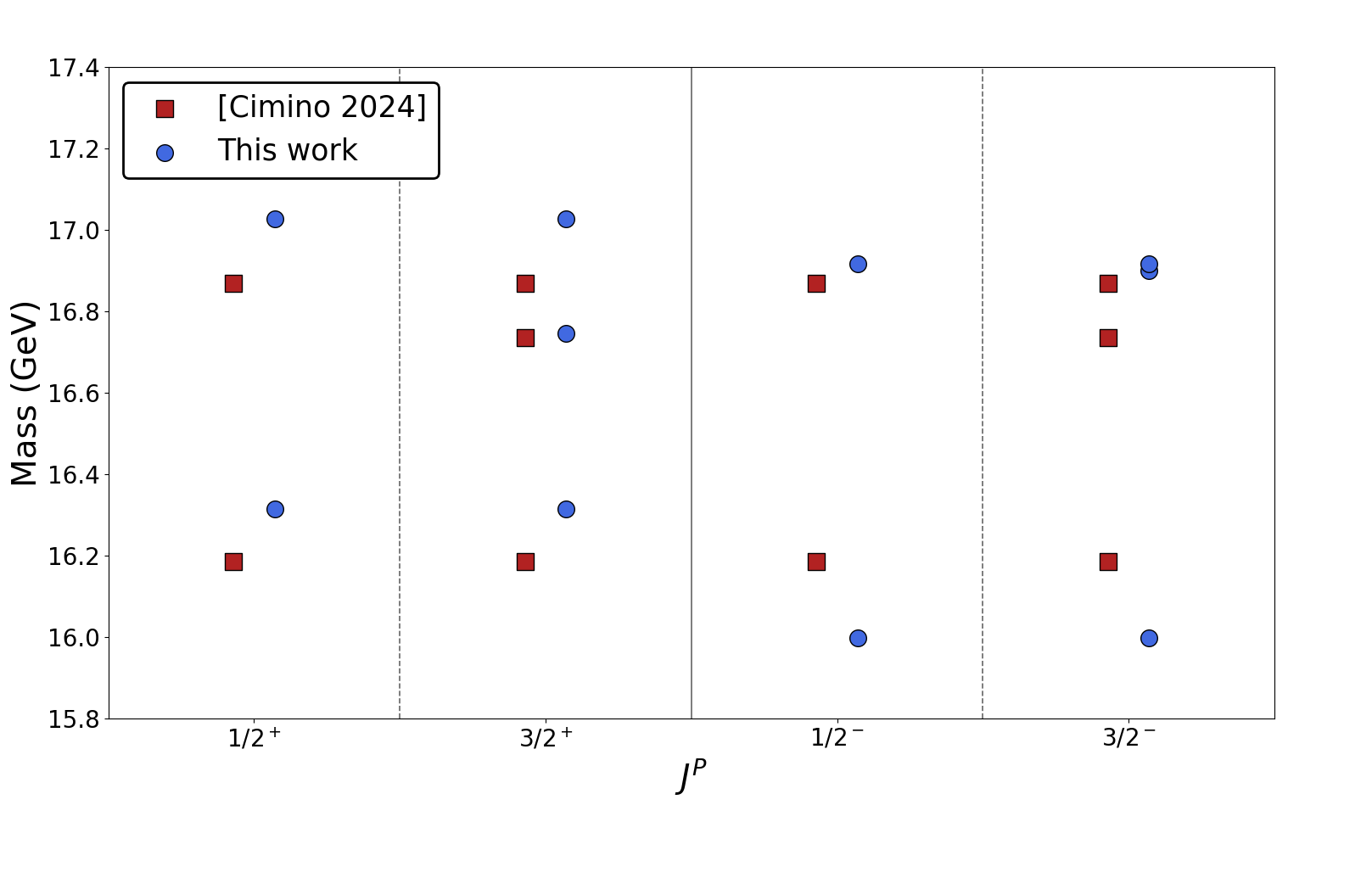}
        \caption{$bbbg$ mass spectrum}
        \label{fig:bbbg}
    \end{subfigure}

    \caption{
        Mass spectrum of heavy hybrid baryons.
        The blue circles correspond to the two or three lowest eigenenergies computed with our model. The red squares correspond to the same eigenenergies computed with the method of Cimino et al.~\cite{Cimi24}.
        }
    \label{fig:Lorenzo vs moi}
\end{figure}

Although the model presented in this work is build for light hybrid baryons, it could be interesting to extend it to the study of heavy ones. The quark core-gluon model being first constructed in Ref.~\cite{Cimi24} for heavy hybrid baryons, and improved and extended in this paper for the light ones, it could be interesting to compare the mass spectrum of the heavy sector calculated with both works. The mass spectra computed for the hybrid baryons $cccg$ and $bbbg$ with the method developed in this paper are provided in Table~\ref{tab:Heavy HB spectrum} for the quark core having quantum numbers $J_C = S_C=1/2$ and $L_C=0$. Note that the masses of the quark-core were obtained via the procedure of Section~\ref{sec:quark core spectrum}, with the mass parameters of the heavy quarks $c$ and $b$ taken as $m_c=1.870$ GeV and $m_b=5.259$ GeV~\cite{Bhad81}. The comparison with the results of Ref.~\cite{Cimi24} is depicted in Fig. \ref{fig:Lorenzo vs moi}.
\setlength{\tabcolsep}{10pt} 

\begin{table}[h!]
\centering
\begin{tabular}{ccccc}
\hline \hline 
 & $m_C$ (Gev)& $J^P$ & $M_0$ (GeV)& $M_1$ (GeV)\\
\hline
\multirow{4}{*}{$cccg$} & \multirow{4}{*}{$5.369$}  
& $\frac{1}{2}^-$ & $6.643$ & $7.508$ \\
& & $\frac{1}{2}^+$ & $6.867$ & $7.600$ \\
& & $\frac{3}{2}^-$ & $6.646$ & $7.402$ \\
& & $\frac{3}{2}^+$ & $6.869$ & $7.282$ \\
\hline
\multirow{4}{*}{$bbbg$} & \multirow{4}{*}{$15.063$} 
& $\frac{1}{2}^-$ & $ 15.998$ & $ 16.918$ \\
& & $\frac{1}{2}^+$ & $16.315$ & $ 17.028$ \\
& & $\frac{3}{2}^-$ & $15.999$ &  $16.900$ \\
& & $\frac{3}{2}^+$ & $16.316$& $16.747$ \\
\hline\hline
\end{tabular}
\caption{Same as in Table~\ref{tab:HB spectrum} but for the mass spectrum of heavy hybrid baryons. For all states, $J_C=1/2$ and $I_C=0$.}
\label{tab:Heavy HB spectrum}
\end{table}

The results show qualitative agreement, even though the parameters of the present model are fitted for the light sector. The differences probably stem from a better treatment of the color structure of the quark core and of the computation of the convoluted potential in this study. This means that our model seems to give coherent results for all the baryonic sector.

\newpage

\section{Conclusion and outlook}
\label{sec: conclusion}

In this work, we have employed a phenomenological potential model, referred to as the \textit{quark core–gluon model} and originally introduced in Ref.~\cite{Cimi24}, with the aim of predicting the mass spectrum of light hybrid baryons composed of three identical quarks. Within this framework, a hybrid baryon is described as an interacting four-body system subject to the following constraints: the three quarks interact among themselves inside a quark core through QCD-inspired two-body interactions, while the remaining constituent gluon interacts with this core, forming an effective two-body system.

This description of the internal dynamics of hybrid baryons, reminiscent of the quark–diquark approximation commonly used for conventional baryons~\cite{tour25}, allows the otherwise demanding four-body quantum problem to be reduced to a three-body problem followed by a two-body one. This reduction considerably simplifies the theoretical treatment and, in particular, enables a consistent and technically tractable implementation of the gluon helicity degrees of freedom. To account for the limitations inherent to this approximation, the interaction between the color-octet quark core and the constituent gluon has been constructed so as to incorporate finite-size effects of the core.

Given the phenomenological nature of the model, its parameters have been fixed by fitting to the experimental masses of light baryons and LQCD results for glueballs. In addition, a hyperfine contribution has been included in the interaction potentials, as such effects are known to be non-negligible for systems composed of light constituents. The results obtained within this approach have been compared with LQCD predictions~\cite{Dudek12} as well as with calculations based on QCD sum rules~\cite{Wang26}. Our model systematically predicts the appearance of hybrid baryon states at higher energies than those found in lattice studies and sum-rule analyses, while showing overall better agreement with the latter.

Owing to certain similarities between the present framework and gluelump-like systems, it is conceivable that an improved agreement with LQCD results could be achieved by supplementing the effective two-body potential with a parity-dependent mass-splitting term, as proposed in Ref.~\cite{Buis08}. In any case, despite quantitative differences with other existing approaches, the model presented in this work supports the general conclusion that the lightest hybrid baryonic resonances are expected to occur at energies above $2~\mathrm{GeV}$. A notable advantage of our method is that it yields a complete mass spectrum across all accessible channels.

Having established a realistic constituent description of hybrid baryons, the future directions are multiple:
\begin{itemize}
    \item Further improve the model by incorporating, for instance, polarization effects of the quark core induced by the constituent gluon, in a spirit similar to the polarization of the electron core of atoms by its peripheral electrons~\cite{Quin17}.
\item Further develop the quark core-gluon model by emphasizing its similarities with gluelump-like systems rather than with two-gluon glueball configurations.
    \item Investigate the decay properties of the hybrid baryons, since such observables are of primary importance for experimental efforts aimed at identifying these exotic states.
    \item Extend the present model to baryonic systems involving quarks of different flavors.
\end{itemize}

\section*{Acknowledgements} 

J.V. and C.C. would like to thank the Fonds de la Recherche Scientifique - FNRS for the financial support. This work was also supported by the IISN under Grant Number 4.45.10.08. The authors also thank E. Canivez and C. Tourbez for their careful reading.


\appendix

\renewcommand{\thesection}{Appendix \Alph{section}}

\renewcommand{\theequation}{\Alph{section}.\arabic{equation}}

\section{Building the basis quark core states with good symmetry}
\label{app:quark core states}

The wave functions for the quark core can be decomposed in the following way:
\begin{equation}
    |\psi_C\rangle = |\rho_L\rangle \otimes |\chi_S\rangle \otimes|\xi_F\rangle\otimes|\phi_c\rangle.
    \label{total WF}
\end{equation}
The states $|\rho_L\rangle,|\chi_S\rangle,|\xi_F\rangle,|\phi_c\rangle $ represent the spatial, spin, flavor and color wave functions respectively. Note that, since we are dealing with three identical fermions,  \eqref{total WF} has to be antisymmetric under exchange of two quarks in the core due to Pauli principle. However, it is not the case with this tensor product decomposition. Therefore, we need to find proper combinations of states which behave well under exchange symmetry. Let us start with the spatial part. In this study, the core spectrum and wave functions are computed via the OBE method~\cite{Chev24}. In this framework, the state $|\rho_L\rangle$ is decomposed in a linear combination of the harmonic oscillator HO basis states in the following way:
\begin{equation}
    \rho_{LM}(\boldsymbol{x},\boldsymbol{y}) = \langle\boldsymbol{x},\boldsymbol{y}|\rho_{LM}\rangle = \sum_{n_x l_x n_y l_y} C_{n_xl_xn_yl_y} \varphi_{n_xl_xn_yl_y}^{LM}(\boldsymbol{x},\boldsymbol{y}),
    \label{app eq: spatial wf}
\end{equation}
where $\bm{x}$ and $\bm{y}$ are internal variables defined in~\cite{Chev24} as dimensionless Jacobi coordinates:
\begin{equation}
    a\bm{x} = \bm{r}_1 - \bm{r}_2, \hspace{4em} \frac{\sqrt{3}a}{2}\bm{y}  = \frac{m_1 \bm{r}_1+m_1\bm{r}_2}{m_1+m_2}-\bm{r}_3,
\end{equation}
with $a$ a variational parameter, and $\bm{r}_1,\bm{r}_2$ and $\bm{r}_3$ the positions of the three quarks in an inertial reference frame. In~\eqref{app eq: spatial wf}, the functions $\varphi_{n_xl_xn_yl_y}^{LM}(\boldsymbol{x},\boldsymbol{y})$ are the two-body HO functions, given by
\begin{equation}
\begin{split}
      \varphi_{n_xl_xn_yl_y}^{LM}(\boldsymbol{x},\boldsymbol{y}) &= [\varphi_{n_xl_x}(\boldsymbol{x})\varphi_{n_yl_y}(\boldsymbol{y})]_{LM} \\
      & = \sum_{m_x,m_y} ( l_x m_x l_y m_y | LM ) \varphi_{n_x l_x m_x}( \boldsymbol{x}) \varphi_{n_y l_y m_y}(\boldsymbol{y}),
\end{split}
\end{equation}
where
\begin{equation}
    \varphi_{n l m}(x,\theta_x,\phi_x) = \mathcal{R}_{nl}(x)Y_{lm}(\theta_x,\phi_x)
\end{equation}
is the solution of the three dimensional harmonic oscillator in spherical coordinates~\cite{Chev24}. The constants $C_{n_xl_xn_yl_y}$ are the expansion coefficients that have to be found. The OBE method is described in detail in Ref.~\cite{Chev24}, but the only important property that we need to keep in mind here is the behavior of the HO basis states under exchange symmetry. For particle $1$ and $2$, it is given by
\begin{equation}
    \mathbb{P}_{12} \varphi_{n_xl_xn_yl_y}^{LM}(\boldsymbol{x},\boldsymbol{y}) = (-1)^{l_x} \varphi_{n_xl_xn_yl_y}^{LM}(\boldsymbol{x},\boldsymbol{y}),
    \label{app eq: spatial symmetry}
\end{equation}
where the operator $ \mathbb{P}_{12} $ encodes the permutation of particle $1$ and $2$. 
For particle $2$ and $3$, it is much more subtle because the action of the corresponding exchange operator, $\mathbb{P}_{23}$, on the Jacobi coordinates mixes them together. This has for consequences the mixing of the HO basis states with same quantum number $L$ under the action of $\mathbb{P}_{23}$. The mixing coefficients can be found through the following relation \cite{Chev24}
\begin{equation}
    \bra{\varphi_{n'_xl'_xn'_yl'_y}^{L'M}(\boldsymbol{x},\boldsymbol{y})} \mathbb{P}_{23}\ket{\varphi_{n_xl_xn_yl_y}^{LM}(\boldsymbol{x},\boldsymbol{y})} = \delta_{L'L}(-1)^{l'_x+l'_y+L}\langle n'_yl'_yn'_xl'_x;L'|n_xl_xn_yl_y;L\rangle_{\pi/6},
    \label{app: brody}
\end{equation}
where $\langle n'_yl'_yn'_xl'_x;L'|n_xl_xn_yl_y;L\rangle_{\pi/6}$ is a Brody-Moshinsky coefficient~\cite{Brod67}.\\

Let us then turn to the study of the color wave function under exchange symmetry. If we were dealing with conventional baryons, $|\phi_c\rangle$ would have been in the color singlet representation, which is completely antisymmetric. But in this work, we are dealing with color octet quark cores. As discussed in Section~\ref{sec:quark core-gluon}, there are two distinct irreducible mixed symmetric eight-dimensional subspaces (see~\eqref{irrep of tensor product of 3 rep}). The first one is mixed antisymmetric, which means that its states are antisymmetric under exchange of the first two particles. On the contrary, the second one is  mixed symmetric, which means that its states are symmetric under the exchange of the first two particles. As for the spatial part, the symmetry under permutation of particles $2$ and $3$ is less obvious. The operator $\mathbb{P}_{23}$ actually mixes the states of the two subspaces together. The mixing coefficients are obtained by expanding the states with a fixed mixed symmetry on particles $1$ and $2$ in terms of those with mixed symmetry on particles $2$ and $3$. This leads to the following mixing matrix
\begin{equation}
    \begin{pmatrix}
        \braket{\phi_c^{MS}|\mathbb{P}_{23}|\phi_c^{MS}} & \braket{\phi_c^{MS}|\mathbb{P}_{23}|\phi_c^{MA}} \\
        \braket{\phi_c^{MA}|\mathbb{P}_{23}|\phi_c^{MS}} & \braket{\phi_c^{MA}|\mathbb{P}_{23}|\phi_c^{MA}}
    \end{pmatrix} = 
    \begin{pmatrix}
        -\frac{1}{2} & \frac{\sqrt{3}}{2} \\ \frac{\sqrt{3}}{2} & \frac{1}{2}
    \end{pmatrix}.
\end{equation}
This will have to be taken into account while building the total three-body wave function.\\

We now turn to the study of the spin and isospin wave functions under exchange symmetry. Since we are considering here an $SU_f(2)$ isospin symmetry for the light quarks $u$ and $d$ only, the treatment of these two wave functions is exactly the same. The irreducible decomposition of the tensor product of three fundamental representations of $SU(2)$ leads to~\cite{Clos79}
\begin{equation}
\begin{split}
\boldsymbol{2}\otimes\boldsymbol{2}\otimes\boldsymbol{2} = \boldsymbol{4}_S \oplus \boldsymbol{2}_{MA}\oplus \boldsymbol{2}_{MS}.
\end{split}
\end{equation}
We see that there is, again, two distinct mixed symmetric subspaces of dimension $2$. The operator $\mathbb{P}_{23}$ then mixes the states of these subspaces together, with mixing coefficients given by~\cite{vars88,silv20}
\begin{equation}
    \braket{\chi_S^{i}|\mathbb{P}_{23}|\chi_S^{j}} = (-1)^{2s_3+s_{12i}+s_{12j}} \sqrt{2s_{12i}+1}\sqrt{2s_{12j}+1} \left\{
\begin{array}{ccc}
s_1 & s_2 & s_{12i} \\
s_3 & S_C & s_{12j}
\end{array}
\right\},
\end{equation}
where $s_k$ is the spin quantum number of the $k$th quark, $i,j$ span the mixed symmetric subspaces, $s_{12i}$ is the spin quantum number of the coupling spin states of particles $1$ and $2$, and $S_C$ is the total spin of the core.\\

Putting all these information together, the total quark core wave function can be obtained in a two-steps procedure:
\begin{itemize}
    \item Select only the basis states with good symmetry under $\mathbb{P}_{12}$ for the spatial, spin, isospin and color parts.
    \item Combine the latter so that they have the same good symmetry under $\mathbb{P}_{23}$ now for each of the spatial, spin, isospin and color parts again.
\end{itemize}
There is, in principle, an infinite number of basis states, which forces us to truncate the expansion. In the OBE method presented in Ref.~\cite{Chev24}, for a given total orbital quantum number $L$, the truncation is performed by selecting only the basis states whose sum $ 2 n_x +l_x +2 n_y +l_y$ is smaller or equal to a fixed even number, noted $Q_{max}$.\\

Let us give an example in order to render this construction clearer. Let us consider $L=0,S=1/2,I=1/2$ and $Q_{max}=0$. The first step is to select the basis states $\ket{\varphi_{n_xl_xn_yl_y}^{LM}(\boldsymbol{x},\boldsymbol{y})}\otimes |\chi_S\rangle \otimes|\xi_F\rangle\otimes|\phi_c\rangle$ which satisfy 
\begin{equation}
\begin{split}
      \mathbb{P}_{12}(\ket{\varphi_{n_xl_xn_yl_y}^{LM}(\boldsymbol{x},\boldsymbol{y})}\otimes |\chi_S\rangle \otimes|\xi_F\rangle\otimes|\phi_c\rangle) &= -  \ket{\varphi_{n_xl_xn_yl_y}^{LM}(\boldsymbol{x},\boldsymbol{y})}\otimes |\chi_S\rangle \otimes|\xi_F\rangle\otimes|\phi_c\rangle\\
      L = l_x + l_y &=0\\
      2n_x + l_x +2n_y+l_y &= 0.
\end{split}
\label{app eq: constraints}
\end{equation}
The two last constraints force all the HO quantum numbers to be zero, which implies the spatial part of the basis states to be symmetric through~\eqref{app eq: spatial symmetry} and~\eqref{app: brody}. Table~\ref{tab:sym_under_P12} provides all the possible basis states that we can construct so that they satisfy the first constraint of~\eqref{app eq: constraints}. Note that, for convenience, these states are arbitrary denoted $\ket{i}$. 
\begin{table}[t]
    \centering
    \begin{tabular}{c c c c}
        \hline\hline
        State & Spin & Isospin & Color \\
        \hline
        $\ket{1}$ & MA & MA & MA \\
        $\ket{2}$ & MA & MS & MS \\
        $\ket{3}$ & MS & MA & MS \\
        $\ket{4}$ & MS & MS & MA \\
        \hline\hline
    \end{tabular}
    \caption{Harmonic-oscillator basis states satisfying the symmetry constraints~\eqref{app eq: constraints}. For convenience, states are arbitrary denoted $\ket{i}$.}
    \label{tab:sym_under_P12}
\end{table}

Any linear combination of the states given in Table~\ref{tab:sym_under_P12}, denoted $\ket{\psi_i}$, will also satisfy the constraints~\eqref{app eq: constraints}. The second step is to construct all the combinations $\ket{\psi_i}$ that are orthogonal and that satisfy the following constraint
\begin{equation}
    \mathbb{P}_{23}\ket{\psi_i} = -\ket{\psi_i}.
    \label{app eq: constraints 2.0}
\end{equation}
In practice, we compute the matrix elements $\bra{j}\mathbb{P}_{23}\ket{k}$, for $i,k\in \{1,2,3,4\}$, diagonalize the resulting matrix, and select only the eigenstates with eigenvalue $-1$. In this example, there is only one eigenstate, given by
\begin{equation}
    \ket{\psi_1} = \frac{1}{2}\ket{1}-\frac{1}{2}\ket{2}-\frac{1}{2}\ket{3}-\frac{1}{2}\ket{4}.
\end{equation}
Therefore, for the truncation $Q_{max}=0$, with $L=0,S=1/2$ and $I=1/2$, there is only one basis state that we can construct with good symmetry properties. This will of course lead to very approximate results. Precise results arise in general for $Q_{max}=8$. 

\section{Computation of the Hamiltonian matrix elements for light hybrid baryons }
\label{app:HB states}

In this section, the Hamiltonian matrix elements of the eigenvalue problem \eqref{eigenvalue problem} are computed. To this end, we have to determine the action of the Hamiltonian operator \eqref{HB hamiltonian} on the hybrid baryon states \eqref{HB trial states}. This operator containing both angular and spin operators, the helicity trial states did not prove to form a suitable basis for the computation. Instead, Ref. \cite{chev25.1} uses a so-called \emph{canonical basis} also composed with states with definite total angular momentum $J,M$
\begin{equation}
    \ket{p,JM,ls;J_C1} = \sum_{m_l,m_s,M_C,\lambda_g} (l m_l s m_s|JM)(J_CM_{C}1\lambda_g |sm_s)\int d \cos \theta d \phi Y_{m_l}^l(\theta,\phi)\ket{p\theta\phi;J_CM_{C}1\lambda_g},
    \label{app eq: canonical basis}
\end{equation}
where the states $|p\theta\phi,J_CM_C1\lambda_g\rangle$ are defined as eigenstates of the relative momentum, core total spin, its projection, the gluon total spin and its projection operators. The associated eigenvalues are given by the quantum numbers  $p, \theta, \phi,J_C, M_C, 1$ and $\lambda_g$. The canonical basis defined in~\eqref{app eq: canonical basis} is built so that its states are eigenstates of the operators $\bm{L}^2$ and $\bm{S}^2$ with associated quantum numbers given by $l$ and $s$, respectively. 
It is possible to relate the two basis, through the following relation \cite{chev25.1,Cimi24}
\begin{equation}
    |p,JM,\lambda_C\lambda_g;J_C\rangle = \sum_{s=|J_C-1|}^{J_C+1} \sum_{l = |J-s|}^{J+s} \sqrt{\frac{2l+1}{2J+1}} (J_C\lambda_C1-\lambda_g|s\lambda_C-\lambda_g)(l0s\lambda_c-\lambda_g|J\lambda_c-\lambda_g)|p,JM,ls;J_C1\rangle.
    \label{app eq: helicity vs canonical}
\end{equation}
Then, by plugging~\eqref{app eq: helicity vs canonical} in~\eqref{HB trial states}, the evaluation of the matrix elements in~\eqref{eigenvalue problem} reduces to a coupled channel analysis with fixed coefficients:
\begin{equation}
\begin{split}
         \langle f_j,J^{P_i},\lambda_C&\lambda_g;J_C|\mathcal{H}_{HB}|f_k,J^{P_i},\lambda'_C\lambda_g;J_C\rangle = \sum_{s=|J_C-1|}^{J_C+1} \sum_{l = |J-s|}^{J+s} \sum_{s'=|J_C-1|}^{J_C+1} \sum_{l' = |J-s'|}^{J+s'}C_{ls\lambda_C\lambda_g}C_{l's'\lambda'_C\lambda_g}\\
&\int\frac{p^2 dp}{2 \sqrt{w_C(p)w_g(p)}}\frac{f_j(p/h)}{p\sqrt{h}} \int\frac{p'^2 dp'}{2 \sqrt{w_C(p')w_g(p')}}\frac{f_j(p'/h)}{p'\sqrt{h}} \bra{p,JM,ls;J_C\lambda_g}\mathcal{H}_{HB}|p',JM,l's';J_C\lambda_g\rangle,
\end{split}
\end{equation}
where
\begin{equation}
    C_{ls\lambda_C\lambda_g} = \sqrt{\frac{2l+1}{2J+1}} (J_C\lambda_C1-\lambda_g|s\lambda_C-\lambda_g)(l0s\lambda_c-\lambda_g|J\lambda_c-\lambda_g).
\end{equation}
The computation of the second line of the equation can be performed using an procedure analogous to~\cite{Chev25.2}, and the generalization to coupled channel is described in~\cite{baye15}.


\bibliographystyle{apsrev4-2}
\bibliography{ref}


\end{document}